\documentclass[12pt]{article}
\usepackage{amsmath}
\usepackage{epsfig}
\begin{document}
\title{Asymptotic behavior of photoionization cross section in a central field}
\author{E. G. Drukarev, A. I. Mikhailov\\
{\em National Research Center "Kurchatov Institute"}\\
{\em B. P. Konstantinov Petersburg Nuclear Physics Institute}\\
{\em Gatchina, St. Petersburg 188300, Russia}}

\date{}
\maketitle

\begin{abstract}
{We demonstrate that the high energy nonrelativistic asymptotics for the photoionization cross section in a central field  $V(r)$ can be obtained without solving of the wave equations for the bound and outgoing electrons. The asymptotics is expressed in terms of the asymptotics of the Fourier transform $V(p)$ of the field. We show that the cross sections drop in the same way for the fields with the Coulomb short distance behavior. The character of the cross sections energy behavior is related to the analytical properties of the function $V(r)$.
The cross sections exhibit power drop for the potentials which have singularities an the real axis. They experience the exponential drop if $V(r)$ has singularities in the complex plane.}
\end{abstract}

\section{Introduction}
In the present paper we study the high energy photoionization. In this process a bound electron is moved to continuum by the photon impact.
In this paper we calculate the asymptotics for the photoionization cross sections without solving the wave equations for the bound and continuum electrons. We assume the electrons to be bound by a local central field $V(r)$ and present the asymptotics in terms of its Fourier transform.  We limit ourselves to the case when the bound electron is in $s$ state. We consider the photon energy $\omega$ which is much larger than the ionization potential $I$ and find the leading term of expansion of the cross section $\sigma(\omega)$ in terms of $1/\omega$. We assume that the photon energy is much smaller than the electron rest energy $m$ (we employ the system of units with $\hbar=1$, $c=1$). Under this limitation the photoelectrons can be treated in nonrelativistic approximation. Hence we analyze the high energy nonrelativistic asymptotics.

The energy of photoelectron is
\begin{equation}
\varepsilon=\omega-I=\frac{p^2}{2m},
\label{1}
\end{equation}
with $p$ the photoelectron momentum. Due to the condition
\begin{equation}
\omega \gg I,
\label{2}
\end{equation}
we can write
\begin{equation}
p \gg \mu,
\label{3}
\end{equation}
where $\mu$ is the characteristic momentum of the bound state. We demonstrate that in asymptotics large momentum $q \approx p \gg \mu$ is transferred to the source of the field mostly by the bound state electron. Thus the energy dependence of the amplitude is determined by the bound state wave function in momentum space
\begin{equation}
\tilde \psi(p)=\int d^3r \psi(r)e^{-i{\bf p}{\bf r}},
\label{4}
\end{equation}
at large $p \gg \mu$. Here $\psi(r)$ is the solution of the wave equation
\begin{equation}
\Big(-\Delta/2m+V(r)\Big)\psi(r)=\varepsilon_B\psi(r); \quad
\varepsilon_B=-I.
\label{4a}
\end{equation}

On the other hand, the Lippmann--Schwinger equation \cite{1} enables to present the wave function $\tilde \psi(p)$
in terms of the Fourier transform of the potential $V(r)$ \cite{2}, which is
\begin{equation}
\tilde V(p)=\int d^3r V(r)e^{-i{\bf p}{\bf r}},
\label{5}
\end{equation}
Thus the asymptotics of the photoionization cross section can be expressed through the potential $\tilde V(p)$.
Since the central field does not depend on direction of ${\bf r}$, we carry out the angular integration and write
\begin{equation}
\tilde V(p)=\frac{4\pi}{p}\int _0^{\infty} dr rV(r)\sin{pr},
\label{5a}
\end{equation}
From now we omit the tilde sign for the functions in momentum space, writing, e.g., $\tilde V(p)$ as $V(p)$.
The explicit expression for the cross section
\begin{equation}
\sigma(\omega)=\frac{4\alpha}{3}\frac{p}{\omega^2}|V(p)|^2\psi^2(r=0); \quad p=\sqrt{2m\omega}.
\label{6}
\end{equation}
will be derived in next Section.
Thus one can find the asymptotical energy dependence of the photoionization cross section without solving the wave equation (\ref{4a}).

Of course, to obtain quantitative values for the cross sections one needs the value of $\psi(r=0)$. However,Eq.(6) predicts the value of the cross sections ratio at two large values of the photon energy
\begin{equation}
\frac{\sigma(\omega_1)}{\sigma(\omega_2)}=\Big(\frac{\omega_2}{\omega_1}\Big)^{3/2}\frac{|V(p_1)|^2}{|V(p_2)|^2}; \quad p_i=\sqrt{2m\omega_i}.
\label{6a}
\end{equation}

In Sec. 2 we derive Eq.(\ref{6}). In Sec. 3 we demonstrate that this equation reproduces the well known asymptotics of the photoionization cross sections in the Coulomb field \cite{3} and in the Dirac bubble potential \cite{4}. We do it just to show that our approach works. In Sec. 4 we consider the potentials with the Coulomb short range behavior, i.e. the potentials with the pole at $r=0$. We show that the photoionization cross section drops as $\omega^{-7/2}$ while the leading corrections compose the Stobbe factor \cite{3}, \cite{2}. In Sec. 5 we  demonstrate that the character of the energy behavior depends crucially on the analytical properties of the potential $V(p)$. The singularities of the function $V(p)$ on the real axis lead to the power drop of the cross section. If $V(p)$
has singularities in the complex plane being  regular on the real axis, we come to exponential drop of the cross section in asymptotics.
We summarize in Sec.6.

Note that Eq.(\ref{6}) and the other  ones are written for the cross sections "per one electron". The physical cross sections should be multiplied by the number $k$ of $s$ electrons in the ionized system. Say, $k=2$ for each atomic shell, $k=1$ for the fullerene ion $C_N^{-}$, etc.

\section{Equation for the asymptotic cross section}

The general expression for the photoionization cross section can be presented as
\begin{equation}
d\sigma=\frac{mp}{(2\pi)^2}|F|^2d\Omega,
\label{7}
\end{equation}
with $F$ being the amplitude of the process. Here averaging over the directions of the photon polarization is carried out.
Now we calculate the amplitude. In photoionization the large momentum ${\bf q}={\bf k}-{\bf p}$ is transferred to the source of the field (here ${\bf k}$ with $k=|{\bf k}|=\omega$ is the photon momentum). One can see that $p \gg k$ except the region near threshold, and thus we can put $q=p$. The recoil momentum $p$ can be transferred either by the bound electron or by the photoelectron. In the former case
the outgoing electron is described by the plane wave, and the amplitude can be written as \cite{2}
\begin{equation}
F_a=N(\omega)\frac{{\bf e}\cdot {\bf p}}{m}\psi(p); \quad N(\omega)=\Big(\frac{4\pi\alpha}{2\omega}\Big)^{1/2}
\label{8}
\end{equation}

Now we employ the Lippman--Schwinger equation written in the momentum space
\begin{equation}
\psi=\psi_0+G(\varepsilon_B)V\psi
\label{9}
\end{equation}
to obtain the function $\psi(p)$ at large values of $p$. Here $G$ is electron propagator of free motion with the matrix elements
$$ \langle {\bf f}_1|G(\varepsilon_B)|{\bf f}_2\rangle=g(\varepsilon_B, f_1)\delta({\bf f}_1-{\bf f}_2); \quad g(\varepsilon_B, f_1)=\frac{1}{\varepsilon_B-f_1^2/2m}.$$
Since for a bound state $\psi_0=0$, we find
\begin{equation}
\psi(p)=\langle {\bf p}|GV|\psi\rangle=g(\varepsilon_B, p)J(p); \quad
J(p)=\int\frac{d^3f}{(2\pi)^3}\langle {\bf p}|V|{\bf f}\rangle\langle{\bf f}|\psi\rangle.
\label{10}
\end{equation}
Due to the bound state wave function $\psi(f) =\langle{\bf f}|\psi\rangle$ the integral $J(p)$ is saturated at $f \sim \mu \ll p$. Thus we can
put $\langle {\bf p}|V|{\bf f}\rangle=\langle {\bf p}|V|0\rangle=V(p)$. Putting also $g(\varepsilon_B, p)=-2m/p^2$ we obtain \cite{2}
\begin{equation}
\psi(p)=-\frac{2m}{p^2}V(p)\psi(r=0)=-\frac{1}{\omega}V(p)\psi(r=0)    .
\label{11}
\end{equation}
Thus Eq.(\ref{8}) can be written as
\begin{equation}
F_a=-2N(\omega)\frac{{\bf e}\cdot {\bf p}}{p^2}V(p)\psi(r=0).
\label{12}
\end{equation}

The outgoing electron also can transfer large recoil momentum. The mechanism is often referred to as the final state interaction. In the lowest order of perturbation theory the corresponding is
$$F_b=N(\omega)\int\frac{d^3f}{(2\pi)^3}\langle {\bf p}|VG(\omega+\varepsilon_B)|{\bf f}\rangle\frac{{\bf e}\cdot {\bf f}}{m}\langle {\bf f}|\psi\rangle$$
\begin{equation}
=N(\omega)\int\frac{d^3f}{(2\pi)^3}\langle {\bf p}|V|{\bf f}\rangle g(\omega+\varepsilon_B, f))\frac{{\bf e}\cdot {\bf f}}{m}\langle {\bf f}|\psi\rangle.
\label{13}
\end{equation}
On the other hand, the amplitude $F_a$ can be written as
\begin{equation}
F_a=
N(\omega)\frac{{\bf e}\cdot {\bf p}}{m}g(\varepsilon_B, p)\int\frac{d^3f}{(2\pi)^3}\langle {\bf p}|V|{\bf f}\rangle\langle {\bf f}|\psi\rangle,
\label{14}
\end{equation}
Since the integral on the right hand side of Eq.(\ref{13}) is saturated by small $f \sim \mu$, we find that $F_b \sim (\mu/p)F_a \ll F_a$.
One can see that the higher order terms in $V$ are quenched by additional powers of $1/p$. Thus the amplitude $F=F_a+F_b$ is dominated by the
first term, and we ca put $F=F_a$ in Eq.(\ref{7}) in calculations of the  asymptotics. Employing Eq. (\ref{12})and carrying out the angular integration we find
\begin{equation}
\sigma(\omega)=\frac{4\alpha p}{3}|\psi(p)|^2.
\label{15}
\end{equation}
Due to Eq.(\ref{11}) this expression is equivalent to Eq.(\ref{6}).

\section{Coulomb and Dirac bubble potentials}

The Fourier transform for the Coulomb field of the point nucleus with the charge $Z$
\begin{equation}
V(r)=-\frac{\alpha Z}{r},
\label{16}
\end{equation}
is $V(p)=-4\pi\alpha Z/p^2$. Employing Eq.(\ref{6})one finds for photoionization of  $ns$ state
\begin{equation}
\sigma=\frac{16\sqrt{2}\pi^2\alpha(\alpha Z)^2}{3m^{3/2}\omega^{7/2}}\psi^2_{ns}(0).
\label{17}
\end{equation}
Recall that in the Coulomb field $\psi_{ns}^2(r=0)=(m\alpha Z)^3/(n^3\pi)$.

The Dirac bubble potential
\begin{equation}
V(r)=V_0\delta(r-R); \quad V_0<0,
\label{18}
\end{equation}
was introduced in \cite{5} and is often used in the fullerene physics. We find immediately
\begin{equation}
V(p)=V_0\frac{4\pi R}{p}\sin{pR}.
\label{19}
\end{equation}
Thus the asymptotic cross section is
\begin{equation}
\sigma=\frac{2^8}{3}\frac{\alpha\pi^2V_0^2m^2R^2}{p^5}\sin^2{(pR)}\psi^2(0); \quad p^2=2m\omega.
\label{20}
\end{equation}

Hence we found the nonrelativistic high energy asymptotics for photoionization in the Coulomb and in the Dirac bubble potentials avoiding the standard procedure in which one should start with solving of the wave equation for the  electron.

\section{Potentials with the Coulomb behavior at short distances}

Here we analyze the large $p$ behavior of Fourier transforms $V(p)$ for the potentials which behave as $1/r$ at $r \rightarrow 0$.

For the Yukawa potential
\begin{equation}
V(r)=-\frac{ge^{-\lambda r}}{r}; \quad g>0,
\label{24}
\end{equation}
we find
\begin{equation}
V(p)=-\frac{4\pi g}{p^2+\lambda^2}.
\label{25}
\end{equation}
At large $p \gg \lambda$
$$ V(p) \approx \frac{-4\pi g}{p^2}(1-\lambda^2/p^2).$$
Thus the asymptotic of the cross section is just the same as in the Coulomb field and is described by Eq.(\ref{17}).
One could expect this result since the asymptotic behavior is determined by the distances $r \sim 1/p \ll 1/\lambda$, where we can put
$e^{-\lambda r}=1$ in Eq.(\ref{24}). We kept the second term in the parenthesis in order to show that the leading corrections to the amplitude $F_a$ are of the relative order $1/p^2$. Hence inclusion of the amplitude $F_b$ providing the correction of the order $1/p$ is more important in analysis of the high energy behavior of the cross section.

The Thomas-Fermi potential in Tietz parametrization \cite{6} is
\begin{equation}
V(r)=-\frac{\alpha Z}{r}\frac{1}{(1+ar/r_0)^2}; \quad a=cZ^{1/3},
\label{26}
\end{equation}
with $c \approx 0.6$ while $r_0=1/m\alpha$ is the Bohr radius. The Fourier transform of this potential is
\begin{equation}
V(p)=-4\pi \alpha ZI(p); \quad I(p)=\frac{1}{p}\int_0^{\infty}dr\frac{\sin{pr}}{( 1+ar/r_0)^2  }=b^2\int_0^{\infty}dr\frac{\cos{pr}}{r+b}.
\label{27}
\end{equation}
with $b=r_0/a$. One can obtain $I$ as a power series of $z=bp$. Including only the two lowest terms, we find
\begin{equation}
I(p)=\frac{1}{p^2}(1-\frac{6a^2}{p^2r_0^2}).
\label{28}
\end{equation}
Thus the asymptotic of $V(p)$ is just the same as in the Coulomb field with the charge $Z$. The leading corrections are of the relative order $1/p^2$.

This is the common feature of the potentials with the Coulomb small distance behavior  $V(r\rightarrow 0) \rightarrow -\alpha Z/r$ approximated by analytical functions at $r>0$.
Since $V(p)$ is determined by small $r$ for large $p$, we can use Eq.(\ref{5}) employing the expansion
\begin{equation}
V(r)=\frac{-\alpha Z}{r}(1+c_1r+c_2r^2)
\label{29}
\end{equation}
We present $V(p)=-\alpha Z[V_0(p)+V_1(p)+V_2(p)]$ with the three terms on the right hand side corresponding to those on the right hand side of Eq.(\ref{29}).
One finds
\begin{equation}
V_0(p)=lim|_{\kappa \rightarrow 0}\int d^3r \frac{e^{-\kappa r}}{r}e^{-i{\bf p}{\bf r}}=\frac{4\pi}{p^2+\kappa^2}|_{\kappa=0};
\label{30}
\end{equation}
$$ V_k(p)=lim|_{\kappa \rightarrow 0}\int d^3r r^k\frac{e^{-\kappa r}}{r}e^{-i{\bf p}{\bf r}}=(-\frac{\partial}{\partial \kappa})^k\frac{4\pi}{p^2+\kappa^2}|_{\kappa=0}. $$
Thus $V_0=4\pi/p^2$, $V_1=0$, $V_2=-8\pi c_2/p^4$, and corrections to the asymptotic term are indeed of the relative order $1/p^2$.
The cross sections are given by Eq.(\ref{17}).

Now we can trace the corrections of the order $1/p$ to the asymptotic law.
As we have seen, such corrections can be provided by the amplitude $F_b$ expressed by Eq.(\ref{13}). Since the integral on the right hand side is saturated by $f \ll p$, we can put $g(\omega+\varepsilon_b, f)=g(\omega,0)$, neglecting thus the terms of the order $f^2/p^2$. We come to the expression
\begin{equation}
F_b=
\frac{N(\omega)}{\omega}\int\frac{d^3f}{(2\pi)^3}\frac{-4\pi\alpha Z}{({\bf p}-{\bf f})^2}\frac{{\bf e}\cdot {\bf f}}{m}\psi({\bf f}).
\label{31}
\end{equation}
 Note that although $f \ll p$, one can not proceed by expanding the denominator in powers of ${\bf p}{\bf f}$ since this would lead to the divergent integral. One can write
$$ \int\frac{d^3f}{(2\pi)^3}{\bf f}\Phi({\bf f}, {\bf p})=A{\bf p}; \quad A=\frac{1}{p^2}\int\frac{d^3f}{(2\pi)^3}({\bf f}{\bf p})\Phi({\bf f}, {\bf p}),$$
for any function $\Phi({\bf f}, {\bf p})$. Thus we can write
$$F_b=
-\frac{N(\omega)\alpha Z}{\omega}\frac{{\bf e}\cdot {\bf p}}{m}\frac {T(p)}{p^2},$$
with
$$ T(p)=\int\frac{d^3f}{(2\pi)^3}\frac{4\pi{\bf f}{\bf p}}{({\bf p}-{\bf f})^2}\psi({\bf f}).$$
Employing Eq.(\ref{4}) we present
$$ T(p)=\int d^3r\psi(r)i{\bf p}{\bf \nabla_r}t({\bf r}),$$
where
$$t({\bf r})=\int\frac{d^3f}{(2\pi)^3}\frac{4\pi}{({\bf p}-{\bf f})^2}e^{-i{\bf f}{\bf r}}=\frac{e^{-i{\bf p}{\bf r}}}{r}.$$
Integrating by parts and noting that
for $s$ states $\psi({\bf r})=\psi(r)$, i.e. does not depend on the direction of vector ${\bf r}$, we find that
$$T(p)=-\int d^3r\frac{e^{-i{\bf p}{\bf r}}}{r}i{\bf p}{\bf \nabla_r}\psi(r)=-i\int\frac{d^3r}{r}e^{-i{\bf p}{\bf r}}{\bf p}{\bf n}\psi'(r),$$
with ${\bf n}={\bf r}/r$.
Thus we can write
\begin{equation}
T(p)=4\pi(T_1(p)+T_2(p)),
\label{31a}
\end{equation}
with
$$ T_1(p)=-\int_0^{\infty}dr\frac{\sin{pr}}{pr}\psi'(r); \quad T_2(p)=\int_0^{\infty}dr\cos{pr}\psi'(r).$$

Calculate first the contribution $T_1$. The integrand is saturated at $r \sim 1/p$ while the derivative of the bound state function scales at
$r \sim 1/\mu \gg 1/p$. Thus we can put $\psi'(r)=\psi'(0)$ and find
$$ T_1(p)=-\frac{\pi}{2p}\psi'(0).$$
To estimate the contribution $T_2$ we carry out integration by parts providing
$$ T_2(p)=-\frac{1}{p}\int_0^{\infty}dr\sin{pr}\psi''(r)= \frac{1}{p^2}(-\psi''(0)+...).$$
Here the dots denote the terms containing the higher terms of the $1/p$ expansion.
Thus $T_2(p)/T_1(p) \sim 1/p$. Looking for the leading term  we can put
\begin{equation}
T(p)=4\pi T_1(p)=-\frac{2\pi^2}{p}\psi'(0),
\label{31b}
\end{equation}
and thus
$$F_b=
\frac{2\pi^2\alpha ZN(\omega)}{\omega}\frac{{\bf e}\cdot {\bf p}}{mp^3}\psi'(0).$$

The exact solution of the wave equation (\ref{4a}) satisfies the first Kato condition \cite{7}, \cite{2}
\begin{equation}
\psi'(0)=-\eta\psi(0); \quad \eta=m\alpha Z.
\label{32}
\end{equation}
Hence
\begin{equation}
F_b=
-\frac{2\pi^2(\alpha Z)^2N(\omega)}{\omega}\frac{{\bf e}\cdot {\bf p}}{p^3}\psi(0).
\label{33}
\end{equation}

Employing Eqs.(11), (14) (15) and (17) we can write
the amplitude $F_a$ for the field with the Coulomb asymptotic as
$$
F_a=
\frac{4\pi\alpha Z N(\omega)}{\omega}\frac{{\bf e}\cdot {\bf p}}{mp^2}\psi(0).
$$
Thus
\begin{equation}
F=F_a+F_b=F_a(1-\frac{\pi\xi}{2}),
\label{34}
\end{equation}
with $\xi=m\alpha Z/p$. Thus the cross section with inclusion of the lowest order correction beyond the asymptotic can be written as
$$\sigma=\frac{16\sqrt{2}\pi^2\alpha(\alpha Z)^2}{3m^{3/2}\omega^{7/2}}(1-\pi\xi)\psi^2(0),$$
see Eq.(\ref{17}).In the higher order terms of the final state interactions in the field $V(r)$ the photoelectron obtains large momenta $f_i \sim p \gg \mu$. Thus these interactions can be  viewed as the exchanges by virtual photons composing the wave function of the outgoing electron. Hence we can use the Coulomb field result in which the terms depending on the parameter $\pi\xi$ compose the Stobbe factor $S(\xi)=\exp{(-\pi\xi)}$ \cite{3}. Note that $\pi\xi$ is not supposed to be small. Thus we obtain
\begin{equation}
\sigma=\frac{16\sqrt{2}\pi^2\alpha(\alpha Z)^2}{3m^{3/2}\omega^{7/2}}\exp{(-\pi\eta/p)}\psi^2(0)(1+O(p^{-2}); \quad \eta=m\alpha Z
\label{34a}
\end{equation}

\section{Cross section behavior and analytical properties of the potential}

Consider first the potentials $V(r)$ which have singularity on the real axis at certain $R>0$.
The cross section for the Dirac bubble potential was found above.
Now we find the cross section for the well potential. This potential has nonzero values only in the limited interval of the real axis,
e.g. at $0 \leq r \leq R$. If the well has the rectangular form, the potential can be written as
\begin{equation}
V(r)=V_0\theta(r)\theta(R-r); \quad V_0<0
\label{37}
\end{equation}
Direct calculation provides
$$ V(p)=-\frac{4\pi V_0}{p^3}(pR\cos{pR}-\sin{pR}).$$
In the asymptotics $pR \gg 1$ the second term in the parenthesis can be neglected, and we obtain
\begin{equation}
V(p)=-\frac{4\pi V_0R}{p^2}\cos{pR}.
\label{38}
\end{equation}
This provides
\begin{equation}
\sigma=\frac{2^7\alpha\pi^2}{3}\frac{V_0^2R^2m}{\omega p^5}\cos^2{(pR)}\psi^2(0); \quad p^2=2m\omega.
\label{39}
\end{equation}

The Dirac bubble potential and the well potential provide the power drop of the cross sections,. It is $\omega^{-5/2}$ in the former case and $\omega^{-7/2}$ in the latter case. This happens because the behavior of the function $\psi(p)$ depends on the character of singularities of the function $\psi(r)$ at $r\rightarrow R$ \cite{8}. Present
\begin{equation}
\psi(p)=\frac{4\pi}{p}\Big[\int_0^{R_{-}}dr\sin{pr}\psi(r)r+\int_{R_{+}}^{\infty}dr\sin{pr}\psi(r)r\Big]; \quad R_{\pm}=R\pm \delta; \quad \delta \rightarrow 0.
\label{40}
\end{equation}
Assuming that $\psi(r)$ is a continuous function at $r=R$ while its first derivative experience s a jump at this point, and integrating by parts we find
\begin{equation}
\psi(p)=\frac{4\pi}{p^2}\Big[\int_0^{R_{-}}dr\cos{pr}(\psi'(r)r+\psi(r))+\int_{R_{+}}^{\infty}dr\cos{pr}(\psi'(r)r+\psi(r))\Big].
\label{41}
\end{equation}
Next integration by parts provides
\begin{equation}
\psi(p)=\frac{4\pi}{p^3}R\sin{(pR)}[\psi'(R_{-})-\psi'(R_{+})]
\label{42}
\end{equation}
$$-\frac{4\pi}{p^3}\Big[\int_0^{R_{-}}dr\sin{pr}(\psi''(r)r+2\psi'(r))+\int_{R_{+}}^{\infty}dr\sin{pr}(\psi''(r)r+2\psi'(r))\Big]$$
Further integration by parts of the second term demonstrates that it is about $1/p$ times the first one. Hence the leading contribution is provided by from  first term. Thus indeed the Fourier transform of the function $\psi(r)$ is determined by the jump of
the first derivative. In the case of the Dirac bubble potential the first derivative experience a jump \cite{5}, and the amplitude drops as $1/p^3$, leading to the $\omega^{-5/2}$ law for the cross section--see Eq.(\ref{15}).  In the case of the well potential the first derivative $\psi'(r)$ is continuous while the second derivative experience a jump, the first term on the right hand side of Req.(\ref{42}) turns to zero, and one more integration by parts should be carried out.

Consider now a simple potential with singularities in the complex plane
\begin{equation}
V(r)=\frac{V_0}{\pi}\frac{a}{r^2+a^2}; \quad V_0<0; \quad a>0,
\label{43}
\end{equation}
for which  $V(r) \rightarrow V_0\delta(r)$ at $a \rightarrow 0$.
One can present
$$V(p)=\frac{4V_0a}{p}\int _{0}^{\infty}dr\frac{r\sin{pr}}{r^2+a^2}.$$
The value of this integral is well known. We present its calculation in order to demonstrate the origin of the exponential quenching of the cross section.
Since the integrand is an even function of $r$ we can write
\begin{equation}
V(p)=\frac{2V_0a}{p}\int _{-\infty}^{\infty}dr\frac{r\sin{pr}}{r^2+a^2}=X_1-X_2,
\label{44}
\end{equation}
with
\begin{equation}
X_1=\frac{V_0}{ip}\int _{-\infty}^{\infty}dr\frac{re^{ipr}}{r^2+a^2}; \quad X_2=\frac{V_0}{ip}\int _{-\infty}^{\infty}dr\frac{re^{-ipr}}{r^2+a^2}
; \quad p>0.
\label{45}
\end{equation}
The integrals $X_1$ and $X_2$ can be calculated in the complex plane by closing the counter in the upper and lower half-planes correspondingly.
They are determined by the poles at $r=\pm ia$ providing
\begin{equation}
V(p)=\frac{2\pi V_0a}{p}\exp{(-pa)},
\label{46}
\end{equation}
and
\begin{equation}
\sigma=\frac{64\pi^2}{3}\frac{\alpha V_0^2m^2a^2}{p^5}\exp{(-2pa)}\psi^2(0),
\label{47}
\end{equation}

One can make a more general conclusion. The cross sections for photoionization of the system bound by the field with singularities in the complex plane experience the exponential drop with $p$. The power of the exponential factor is twice the imaginary part of the singularity closest to the real axis.

We conclude this Section by considering the case of the modified P$\ddot{o}$schl-Teller potential \cite{9}
\begin{equation}
V(r)=\frac{V_0}{\cosh^2(\kappa r)}; \quad V_0<0; \quad \kappa>0.
\label{48}
\end{equation}
Its Fourier transform is \cite{10}
\begin{equation}
V(p)=\frac{\pi^3V_0}{\kappa^3}\exp{(-\pi p/2\kappa)}.
\label{55}
\end{equation}
Thus the cross section of photoionization is
\begin{equation}
\sigma(\omega)=\frac{4\alpha\pi^6V_0^2}{3\kappa^6}\frac{p}{\omega^2}\exp{(-\pi p/\kappa)}\psi^2(0); \quad p=\sqrt{2m\omega}.
\label{56}
\end{equation}
The cross section experience the exponential drop with $p$ since the potential (\ref{48}) has poles at $r=\pm i\pi(2n+1)/2\kappa$ where $n$ is a natural number.
The asymptotics is determined by the residue ar $r=i\pi/2\kappa$.

\section{Summary}

We demonstrated that the high energy nonrelativistic asymptotics of the photoionization cross section in the central field $V(r)$ can be obtained without solving the wave equation for the bound electron and for the photoelectron. The asymptotics can be expressed in terms of the Fourier transform $V(p)$. The expression for the cross section is given by Eq.(\ref{6}). It reproduces the well known results for photoionization in the Coulomb field and in the Dirac bubble field. 

We found that in any field with the Coulomb behavior at short distances the cross section has the same $\omega^{-7/2}$ drop. The leading corrections to the asymptotic law are given by the universal Stobbe factor. The photoionization cross sections experience the power drop for all the potentials with the singularities at the real axis. The power is determined by the analytical properties of the solutions of the wave equation at the singular point.
In the case of the Dirac bubble potential the first derivative of the wave function $\psi'(r)$ has a jump and the cross section drops as $\omega^{-5/2}$.
In the well potential the first derivative is continuous while the second derivative experience a jump. This provides the $\omega^{-7/2}$ law for the cross section.

 It is demonstrated that the photoionization cross sections exhibit the exponential drop if the function  $V(r)$ is regular on the real axis but has singularities in the complex plane. The statement is illustrated by calculation of the cross sections for the potentials given by Eq.(\ref{43}) and (\ref{48}).


\begin{thebibliography}{}


\bibitem{1}R. Newton, {\em Scattering Theory of Waves and Particles},
Springer-Verlag NY, 1982.
\bibitem{2} E. G. Drukarev and A. I. Mikhailov,{\em High -- Energy Atomic Physics}, Springer International Publishung AG Switzerland 2016.
\bibitem{3} H. A. Bethe and E. E. Salpeter, {\it Quantum Mechanics of One- and Two--Electron Atoms},
Dover Publications, NY, 2008.
\bibitem{4} M. Ya. Amusia, A. S. Baltenkov and B. G. Krakov, Phys. Lett. A {\bf 243}, 99 (1998).
\bibitem{5} L. L. Lohr and S. M. Blinder, Chem. Phys. Lett. {\bf 180}, 100 (1992).
\bibitem{6} T. Tietz, Zs. Naturforsch., {\bf 23a}, 191 (1968).
\bibitem{7} T. Kato, Commun. Pure Appl. Math. {\bf 10}, 151 (1957).
\bibitem{8} A. B. Migdal, {\em Qualitative Methods in Quantum Physics}, Perseus Books, Reading, MA USA, 2000.
\bibitem{9} S. Fl$\ddot{u}$gge, {\em Practical Quantum Mechanics--I}, Springer-Verlag, Berlin--Heilderberg-- NY, 1971.
\bibitem {10}{\em Tables of Integral Transforms},edited by A. Erdelyi, v.1, New York--Toronto--London Mc Graw Hill, 1954.


\end{thebibliography}
\end{document}